\documentstyle[twoside,natbib,epsf]{article}

\input{ibvs2.sty}

\def\cite#1{\citealt{#1}}
\def\ibvs{Inf. Bull. Var. Stars}
\def\aap{A\&A}

\def\apj{ApJ}

\def\baas{BAAS}
\def\mnras{MNRAS}
\def\pasj{PASJ}
\def\pasp{PASP}

\def\inpress{in press}
\def\astroph#1{ (astro-ph/#1)}

\begin{document}

\IBVShead{xxxx}{xx July 2002}

\IBVStitletl{NSV 10934: an Unusual ROSAT-Selected Dwarf Nova:}{An Analogous Object to HT C\lowercase{am}?}

\IBVSauth{Kato,~Taichi$^1$, Stubbings,~Rod$^2$, Yamaoka,~Hitoshi$^3$, Nelson,~Peter$^4$, \\
Monard,~Berto$^5$, Pearce,~Andrew$^6$, Garradd,~Gordon$^7$
}
\vskip 5mm

\IBVSinst{Dept. of Astronomy, Kyoto University, Kyoto 606-8502, Japan, \\
          e-mail: tkato@kusastro.kyoto-u.ac.jp}

\IBVSinst{19 Greenland Drive, Drouin 3818, Victoria, Australia,
          e-mail: stubbo@sympac.com.au}

\IBVSinst{Faculty of Science, Kyushu University, Fukuoka 810-8560, Japan,
          e-mail: yamaoka@rc.kyushu-u.ac.jp}

\IBVSinst{RMB 2493, Ellinbank 3820, Australia,
          e-mail: pnelson@dcsi.net.au}

\IBVSinst{Bronberg Observatory, PO Box 11426, Tiegerpoort 0056, South Africa,
          e-mail: LAGMonar@csir.co.za}

\IBVSinst{32 Monash Ave, Nedlands, WA 6009, Australia,
          e-mail: apearce@ozemail.com.au}

\IBVSinst{PO Box 157, NSW 2340, Australia,
          e-mail: loomberah@ozemail.com.au}

\IBVSobj{NSV 10934}
\IBVStyp{UG}
\IBVSkey{dwarf nova, photometry, astrometry, identification, classification}

\begintext

   NSV 10934 was discovered as a large-amplitude suspected variable star
of unknown classification.  The cataloged range of variability was
11.2 to [15.0 p.  We noticed that the object can be identified with
a bright ROSAT X-ray source (1RXS J184050.3-834305).  Since a combination
of a large-amplitude variation and the strong X-ray emission suggests
a cataclysmic variable, we started systematic monitoring of this
NSV object through the VSNET Collaboration (vsnet-chat 3340).\footnote{
  http://www.kusastro.kyoto-u.ac.jp/vsnet/Mail/chat3000/msg00340.html
}  The first outburst was detected on 2001 March 13 by RS (vsnet-alert
5778).\footnote{
  http://www.kusastro.kyoto-u.ac.jp/vsnet/Mail/alert5000/msg00778.html
}  Three additional outbursts have been recorded since 2002 July.
The well-observed most recent two outbursts have been characterized
by a sudden rise (more than 1.3 mag within 1 d), which established the
dwarf nova-type variability.  Table 1 lists the observed outbursts.
Although the rising phase was not covered by observations, the observed
maximum magnitudes suggest that the first two outbursts were detected
close to their maxima.  Figure 1 depicts the long-term light curve based
on visual observations by the authors (RS, NP, PA) and snapshot CCD
observations.  The accuracy of the visual observations is 0.2--0.3 mag,
which will not affect the following discussion.
The shortest interval of the observed outbursts is 46 d.

   Astrometry of NSV 10934 was performed on CCD images taken by PN
(2002 June 5.423 UT) and BM (2002 July 21.024 UT),
both of which were taken during the rapid decline stage from outbursts.
The variability of the object has been confirmed by a comparison between
the two images.  An average of measurements of two images
(UCAC1 system, 182 and 71 reference stars respectingly; internal
dispersion of the measurements was $\sim$0$''$.1) has yielded
a position of 18$^h$ 40$^m$ 52$^s$.52, $-$83$^{\circ}$ 43$'$ 09$''$.84
(J2000.0).  The position agrees with the USNO$-$A2.0 star at
18$^h$ 40$^m$ 52$^s$.28, $-$83$^{\circ}$ 43$'$ 09$''$.2
(epoch 1983.040 and magnitudes $r$ = 15.6, $b$ = 16.5), or the
GSC-2.2.1 star with position end figures of 52$^s$.420 and 09$''$.74 
(epoch 1993.767 and magnitudes $r$ = 15.11, $b$ = 17.06),
which is most likely the quiescent counterpart of NSV 10934 (Figure 2).
This identification has confirmed the previously proposed
identification by \citet{lop90varastrometry} better than an accuracy
of 1$''$. 

\IBVSfig{6.5cm}{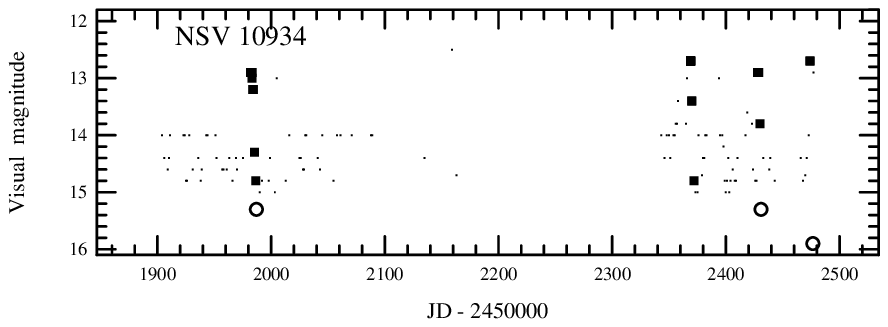}{
   Long-term light curve of NSV 10934.  Large and small dots
represent positive and negative (upper limit) observations, respectively.
Open circles are unfiltered CCD measurements, which have a sensitivity
close to $R_{\rm c}$.  The most recent two CCD observations (NP and NP)
have been calibrated by using GSC 9523.1025 (Tycho-2 magnitude: $V$ = 11.73,
$B-V$ = +1.07).  The overall uncertainty of the CCD photometry is 0.2 mag.
}

\refstepcounter{figure}
\begin{figure}[hpt]
\normalsize
\vskip 3mm plus 1mm minus 1mm
\centerline{\epsfysize=8.5cm \epsffile{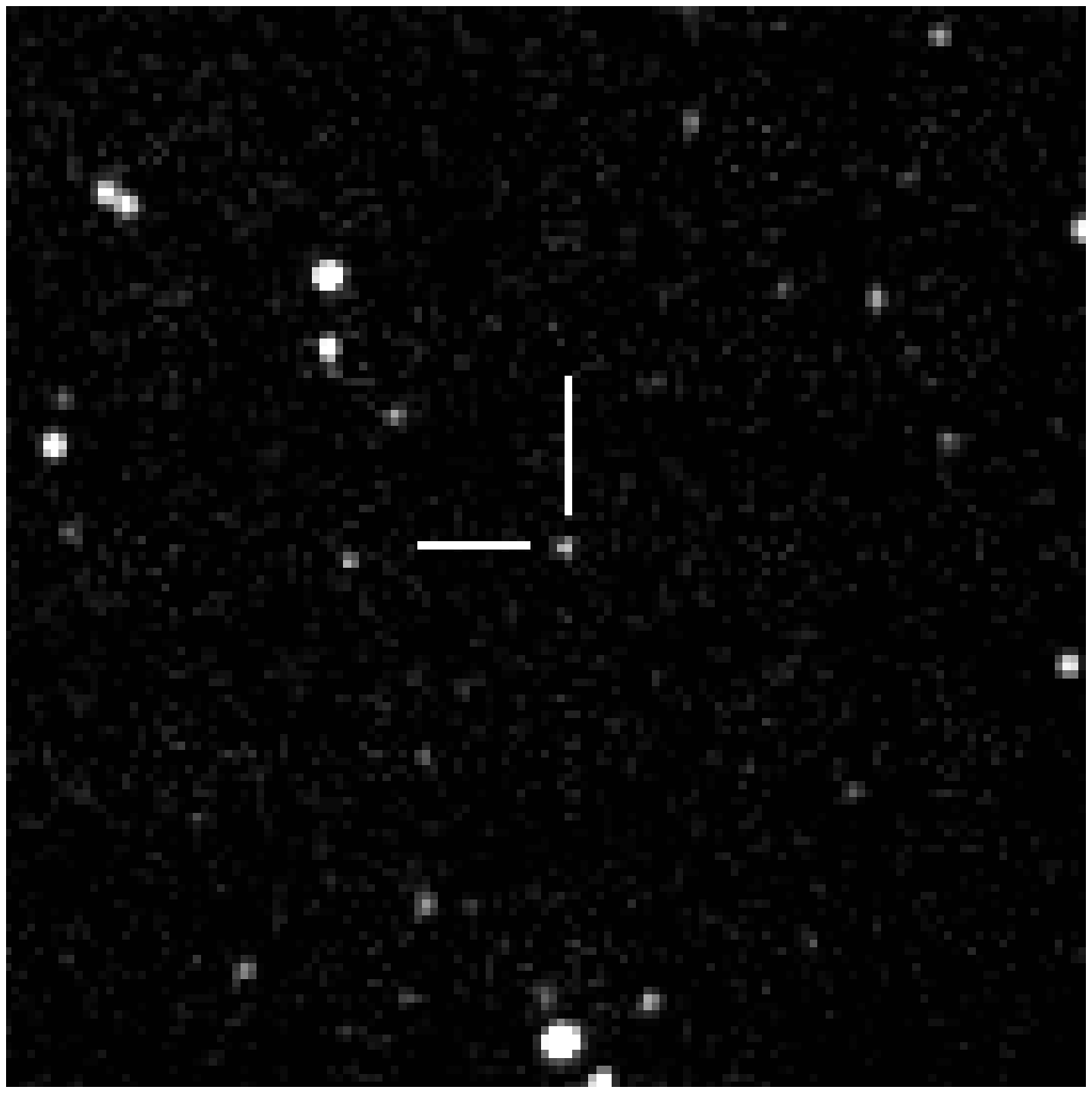} \hskip 2mm
            \epsfysize=8.5cm \epsffile{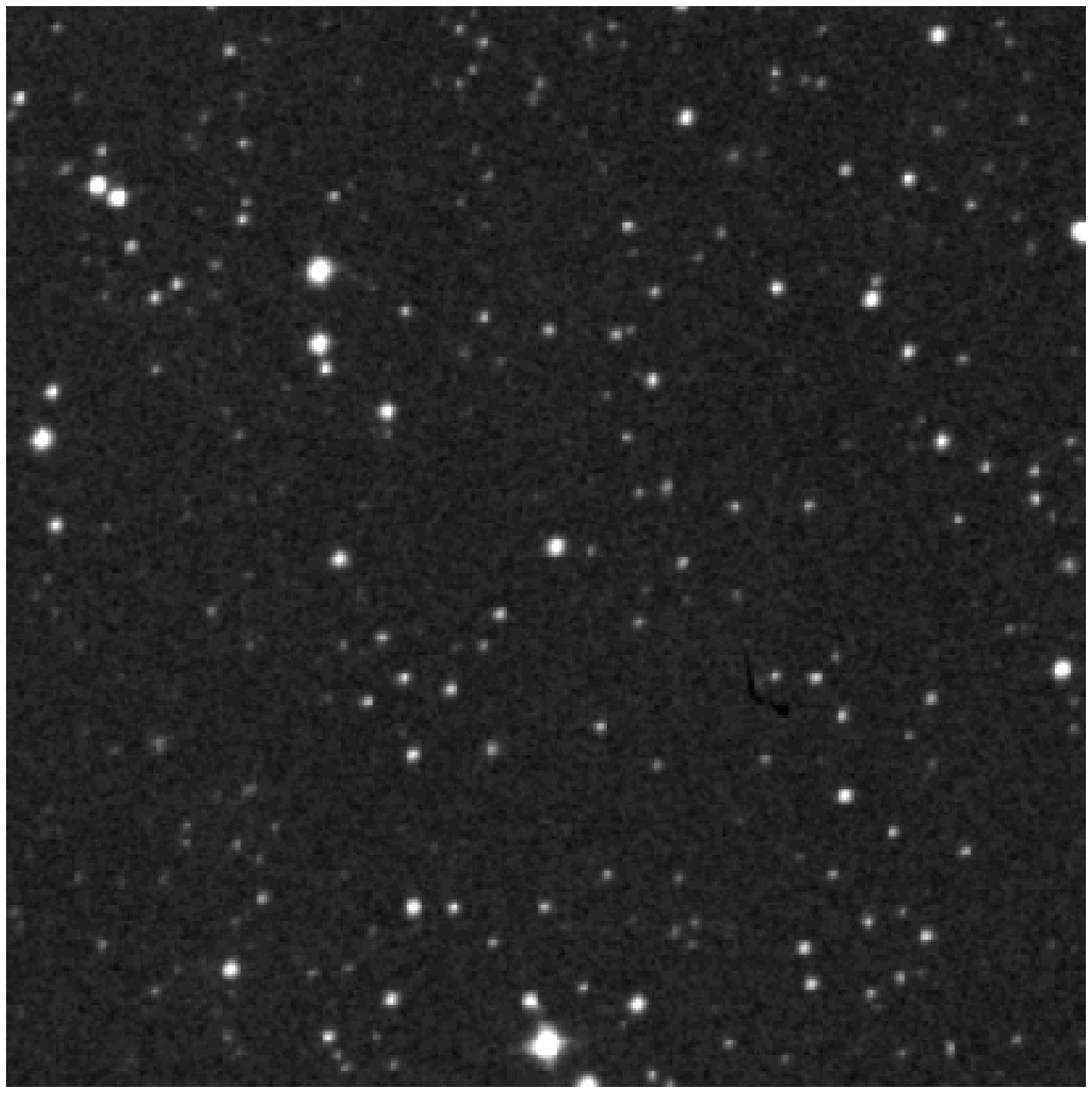}}
\vskip 2mm plus 1mm minus 1mm
\begin{center}
{\figfon Figure \thefigure .} 
\figcap Identification of NSV 10934.  (Left) PN's image on 2002 June 5.423 UT
(8 arcminutes square, north is up, and east is left; magnitude 15.3,
slightly above quiescence).  (Right) DSS red image showing NSV 10934 in
quiescence. \large
\end{center}
\end{figure}
\vskip 4mm plus 1mm minus 1mm

   Figure 3 shows the enlarged light curve of the best observed outbursts.
All the recorded outbursts rather quickly faded.  The most recent two
outbursts faded more than 1 mag within 3 d of the outburst maximum.
Linear fits to the best-observed decline stages of the first two
outbursts have yielded decline rates of 0.71$\pm$0.06 mag d$^{-1}$ and
0.71$\pm$0.03 mag d$^{-1}$, respectively.  Rather fragmentary data of
the recent two outbursts further suggest an even higher value close
to the termination of the outbursts: the object faded by 1.5 mag in 0.90 d
(between JD 2452430.02 and 2452430.92) and by 3.2 mag in 2.59 d (between
JD 2452473.93 and 2452476.52).

\begin{table}
\begin{center}
Table 1. Outbursts of NSV 10934. \\
\vspace{10pt}
\begin{tabular}{ccccc}
\hline
\multicolumn{3}{c}{Date} & JD-2400000 & Max \\
\hline
2001 & March & 13 & 51982 & 12.9 \\
2002 & April & 4  & 52369 & 12.7 \\
2002 & June  & 2  & 52428 & 12.9 \\
2002 & July  & 18 & 52474 & 12.7 \\
\hline
\end{tabular}
\end{center}
\end{table}

\IBVSfig{9cm}{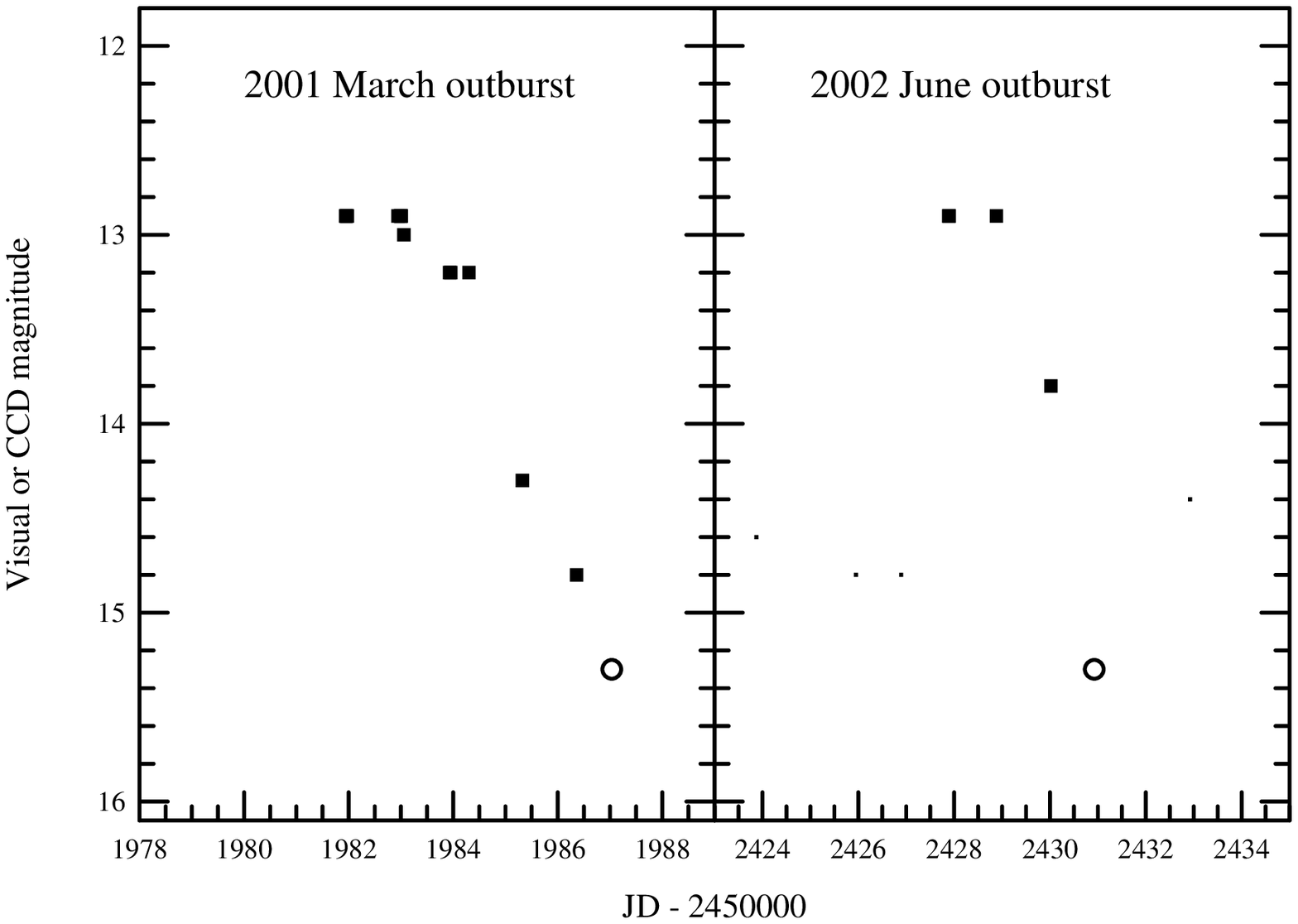}{
  Enlarged light curve of the best observed outbursts of NSV 10934.
The symbols are the same as in Figure 1.
}

   The combination of relatively strong and relatively hard (hardness ratios:
HR1 = 1.00, HR2 = 0.50) X-ray detection and short optical outbursts
either suggests the possibility of an intermediate polar (IP)
with dwarf nova-like outbursts, a non-magnetic HT Cas-like unusual dwarf
nova with rather irregularly spaced short outbursts
\citep{kat02ircom,muk97htcas,woo95htcasXray}, or a system resembling
an unusual dwarf nova BZ UMa with short outbursts and quasi-periodic
oscillations \citep{kat99bzuma,wen82bzuma,rin90bzuma,jur94bzuma}.

\begin{table}
\begin{center}
Table 2. Comparison of X-ray Properties of NSV 10934 and HT Cam$^a$ \\
\vspace{10pt}
\begin{tabular}{ccccc}
\hline
Object    & Count rate & HR1 & HR2 & $V$ \\
\hline
NSV 10934 & 0.239 & 1.00 & 0.50 & 15.9 \\
HT Cam    & 0.152 & 0.79 & 0.43 & 16.2 \\
\hline
 \multicolumn{5}{l}{$^a$ The X-ray data are taken from \citet{ROSATRXP}.} \\
\end{tabular}
\end{center}
\end{table}

   The precipitous fading (1.5 mag in 0.90 d) recorded during the terminal
stage of the outbursts is unlike usual dwarf novae.  The sequence of
a more slowly fading plateau phase near outburst maxima and a subsequent
rapid fading more resembles the behavior of an outburst in the recently
discovered IP, HT Cam \citep{ish02htcam}.  The X-ray hardness ratios are
also similar (Table 2).  HT Cam showed a gradual decline
for the first 0.5 d, followed by a dramatic decline by more than 4
mag d$^{-1}$ \citep{ish02htcam}.  Since the time-evolution of the light
curve is slightly slower in NSV 10934, the orbital period of NSV 10934
is expected to be slightly longer than that of HT Cam (86 min), if
NSV 10934 indeed turns out to be a HT Cam-like object.

   In any case, NSV 10934 is an unusual dwarf nova which deserves further
attention.

\vskip 3mm

This work is partly supported by a grant-in aid [13640239 (TK),
14740131 (HY)] from the Japanese Ministry of Education, Culture, Sports,
Science and Technology.
The CCD operation of the Bronberg Observatory is partly sponsored by
the Center for Backyard Astrophysics.
The CCD operation by Peter Nelson is on loan from the AAVSO,
funded by the Curry Foundation.
This research has made use of the Digitized Sky Survey producted by STScI, 
the ESO Skycat tool, and the VizieR catalogue access tool.

\end{document}